\documentclass[12pt]{article}
\newcommand{\R}{R}
\newcommand{\sect}[1]{\section{#1}\setcounter{equation}{0}}

\newcommand{\OL}[1]{ \hspace{1pt}\overline{\hspace{-1pt}#1
   \hspace{-2pt}}\hspace{2pt} }
\newcommand{\skyp}[1]{}
\newcommand{\half}{{\textstyle \frac{1}{2}}}

\begin{document}

\bigskip
\hskip 5in\vbox{\baselineskip12pt
\hbox{NSF-ITP-02-120}
}
\bigskip\bigskip

\centerline{\Large \bf The Quantum Hall Effect on $R^4$
}
\bigskip
\centerline{\bf Henriette Elvang}
\medskip
\centerline{Department of Physics}
\centerline{University of California}
\centerline{Santa Barbara, CA 93106}
\centerline{\it elvang@physics.ucsb.edu}
\bigskip
\centerline{\bf Joseph Polchinski}
\medskip
\centerline{Institute for Theoretical Physics}
\centerline{University of California}
\centerline{Santa Barbara, CA\ \ 93106-4030}
\centerline{\it joep@itp.ucsb.edu}

\begin{abstract}
Zhang and Hu have formulated an $SU(2)$ quantum Hall system on the
four-sphere, with interesting three-dimensional boundary dynamics including
gapless states of nonzero helicity.  In order to understand the local physics
of their model we study the
$U(1)$ and
$SU(2)$ quantum Hall systems on flat $\R^4$, with flat boundary $\R^3$.  In the
$U(1)$ case the boundary dynamics is essentially one dimensional.  The $SU(2)$
theory can be formulated on $\R^4$ for any isospin $I$, but in order to obtain
a flat boundary theory we must take $I \to \infty$ as in Zhang and Hu.  The
theory simplifies in the limit, the boundary becoming a collection of
one-dimensional systems.  We also discuss general constraints on the emergence
of gravity from nongravitational field theories.

\end{abstract}

\newpage
\baselineskip=18pt

\sect{Introduction}

The two-dimensional quantum Hall effect (QHE) has been a rich and
fascinating subject.  The bulk has a mass gap, and so the
low-lying excitations live on the one-dimensional edge.  Many nontrivial
phenomena of $(1+1)$-dimensional quantum field theory arise in the QHE edge
dynamics.

Recently, Zhang and Hu have found a beautiful four-dimensional
generalization of the QHE, with three-dimensional edge
dynamics, based on fermions moving in a background $SU(2)$ gauge
field~\cite{hz1,hz2}.   Their most striking result is the presence of
gapless spin-two bosons in the edge theory, suggesting the
emergence of gravity.  The model as presently formulated
is a free theory, so there is no gravitational force, and there are
actually massless bosons of all helicities.  However, it has been
argued~\cite{hz1} that introducing interactions might plausibly remove the
unwanted states while leaving a theory of gravity.

Our goal is to develop a better understanding of the {\it local}
dynamics of the Zhang-Hu model, where most of the key
physics issues should arise.
The model is originally formulated with the spatial dimensions forming a
four-sphere $S^4$.  To expose the local physics one must take the
infinite-radius limit while focusing on a patch with geometry $\R^4$.
In the Zhang-Hu model this limit is nontrivial: the fermions couple to
the background gauge field with isospin $I$, and one must take $I$
to infinity along with the radius.  We would like to understand better why
this is necessary, and in what sense the limit exists.  Further, if the
limit does exist then we might hope that it allows for some simplification,
so that the important aspects of the physics are clearer than in the
formulation on $S^4$.

Let us mention in particular one puzzling feature of the Zhang-Hu model.
The `graviton' is a particle-hole state.  It is
argued in refs.~\cite{hz1,hz2} that the particle-hole separation remains
small at all times, even in the absence of interactions, so that one can
think of the state as a single particle.  
However, the uncertainty principle normally forbids this.  If the separation
is initially finite, $|\delta \vec x| < \infty$, then the relative
momentum of the particle and hole is uncertain, $|\delta \vec p| > 0$.  But
the velocity is in general a nontrivial function of the momentum, so that
$|\delta \vec v| > 0$ as well, and then the separation will grow linearly in
time.  The one exception to this is for relativistic particles in one
dimension, which move with velocity $c$ independent of their momentum.
This is the essence of bosonization: a noninteracting fermion-antifermion pair
forms a bosonic excitation that remains localized.  But in more than one
dimension $\partial v^i/\partial p^j$ is nontrivial (in particular the
direction of the velocity depends on that of the momentum), and there is no
natural bosonization.

Our approach will be to formulate the quantum Hall effect directly on flat
$\R^4$, making contact with the Zhang-Hu model only later.  In section 2 we
consider the QHE based on gauge group $U(1)$.  We first review the
two-dimensional theory and its edge dynamics.  We then extend this to four
dimensions in the obvious way, by introducing $U(1)$ magnetic fields in two
independent planes.  We show that the edge dynamics is not truly
three-dimensional.  Rather, it corresponds to a one-dimensional system with
an infinite number of fermion fields, with helicities $0,1,2,\ldots\ $, or
equivalently to parallel one-dimensional systems arrayed (fuzzily) in two
transverse dimensions.  Nevertheless, this system turns out to be a useful
building block toward understanding the $SU(2)$ system.  By taking a particle
and hole with different helicities, we obtain localized gapless
particle-hole excitations of arbitrary helicity as claimed in ref.~\cite{hz1}. 
We develop some of the properties of these states, and we find some curious
aspects that may be an obstacle to a relativistic theory.

The failure of the
$U(1)$ example can be ascribed to insufficient spatial symmetry.  The
symmetry group is $U(2)$, which is smaller than the spatial symmetry group
(rotations plus translations) of
$\R^3$.  In section~3 we show that by introducing an $SU(2)$ gauge field as
in ref.~\cite{hz1}, it is possible to retain an $SO(4)$ symmetry that
combines spatial rotations with gauge rotations.  This reduces to the spatial 
symmetry group of $\R^3$ in the flat limit.  We are able to
formulate, and solve, this version of the QHE on flat $\R^4$ even for
finite isospin
$I$.  However, the
density of states in the lowest Landau level of our system is finite for
finite $I$.
A bubble of quantum Hall fluid thus has a maximum radius, so the
edge theory lives on $S^3$ not $\R^3$. In order to take the limit of a large
bubble of quantum Hall fluid, so that its edge becomes locally $\R^3$, we find
it necessary to take
$I \to \infty$ just as in ref.~\cite{hz1}.\footnote{There is another case in
which the number of lowest Landau level states is infinite but the local
density diverges at large radius, which is also unsatisfactory for going to
the $\R^3$ limit.}

In section 4 we simplify the system to the maximum extent possible by
taking the $I \to \infty$ limit of our system at the beginning, before taking
the size of the Hall bubble to be large.  The result is a continuously
infinite collection of four-dimensional $U(1)$ systems, distinguished by the
spatial orientation of the magnetic field.  The corresponding edge theory is
an infinite collection of one-dimensional theories, distinguished by their
orientation in three dimensions.

Section 5 is somewhat independent from the rest, an essay about emergent
gravity.  We explain why we do not believe that this is possible in the
Zhang-Hu approach, and contrast this with the AdS/CFT duality which is an
example of emergent gravity.  We also relate this to the more familiar
phenomenon of emergent gauge symmetry.

Ref.~\cite{kara} considers both $U(1)$ and $SU(2)$ magnetic fields on $CP^2$,
so the discussion in our section~2.2 would govern the local and edge
dynamics of the $U(1)$ case.  Refs.~\cite{followup} develop the Zhang-Hu
idea in other directions; it may be interesting to consider the local limits of
these.

\sect{The $U(1)$ QHE on $\R^2$ and $\R^4$}

\subsection{The $U(1)$ QHE in two dimensions}

\subsubsection{The bulk}

We first review the physics of charged fermions in a constant magnetic
field in two dimensions.  For simplicity the fermions are spinless.
We use units $\hbar = e/c = 1$, so the covariant derivative is $D_a =
\partial_a - i A_a$.  The spatial
dimensions are indexed $a,b$; since these are spatial indices, there is no
distinction between upper and lower.  We work in the gauge
\begin{equation}
A_1 = -\frac{B}{2} x_2\ ,\quad A_2 = \frac{B}{2} x_1\ .
\label{vectpot}
\end{equation}
The Hamiltonian is
\begin{eqnarray}
H &=& -\frac{1}{2m} D_a D_a \nonumber\\
&=& \frac{1}{2m} \biggl( -\partial_a \partial_a + \frac{B^2}{4} x_a
x_a - B L_{12} \biggr) \nonumber\\
&=& \frac{|B| (n + 1) - B L_{12}}{2m} 
\ .
\end{eqnarray}
Here $n$ is the total number of oscillator excitations and
\begin{equation}
L_{ab} =  -i(x_a \partial_b - x_b \partial_a)\ .
\end{equation}

For $B > 0$ the lowest Landau level (LLL)
consists of all states with $L_{12}=n$; these have the minimum energy
$B/2m$.
It is convenient to work with complex coordinates,
\begin{eqnarray}
z &=& \frac{1}{2}(x_1 + i x_2)\ , 
\quad\partial_z = \partial_1 - i\partial_2\ ,
\nonumber\\
D_z &=& \partial_z - B \bar z\ ,\quad
D_{\bar z} = \partial_{\bar z} + Bz\ .
\end{eqnarray}
The Hamiltonian is then
\begin{equation}
H = \frac{B}{2m} - \frac{1}{2m} D_z D_{\bar z}\ .
\end{equation}
The second term is nonnegative and for
$B > 0$ the LLL states satisfy $D_{\bar z} \psi = 0$, implying that
\begin{equation}
\psi = f(z) \exp(-B z \bar z)
\end{equation}
with $f(z)$ analytic.  The case $B < 0$ is given by $ z
\leftrightarrow \bar z$, so without loss of generality we take $B$ positive
in the remainder of this section.

The system is translationally invariant, and so there exist magnetic
translation operators $\Pi_a$ having the property
\begin{equation}
[\Pi_a,D_b]=0\ .
\end{equation}
In the gauge~(\ref{vectpot}) these are simply given by $\Pi_a =
-i(\partial_a + i A_a)$. There are two convenient bases for the LLL.  The
first are the eigenstates of
$L_{12}$,
\begin{equation}
f(z) \propto z^l\ ,\quad n = L_{12} = l\ .
\end{equation}
The second are the eigenstates of $\Pi_1$,
\begin{equation}
f(z) \propto \exp(B z^2 + 2ip_1 z)\ ,\quad  \Pi_1 = p_1\ .
\end{equation}
In the latter case, $|\psi|$ is independent of $x_1$ and gaussian in $x_2$.

\subsubsection{The edge}

To produce a localized bubble one adds a confining potential to the
Hamiltonian (we also add a constant so that the LLL energy is zero):
\begin{equation}
H' = H - \frac{B}{2m} + V\ , \quad V = \frac{\kappa}{2} x_a x_a\ ,
\label{conf}
\end{equation}
with $\kappa$ a positive constant.
Now take the limit $m \to 0$.  In this limit all excited states go to
infinite energy and so only the LLL states mix under $V$; we can write
\begin{equation}
H' = V \quad (\mbox{between LLL states})\ .
\end{equation}
By
rotational invariance, $V$ is diagonal in the $L_{12}$ basis, and
therefore so is the Hamiltonian
\begin{equation}
\langle l | x_a x_a | l' \rangle = \frac{2}{B} (l+1) \delta_{ll'} \
,\quad
\langle l | H' | l' \rangle = \frac{\kappa}{B} (l+1) \delta_{ll'} \ .
\label{radius}
\end{equation}

The second-quantized Hamiltonian is 
\begin{equation}
{\bf H}' = \frac{\kappa}{B} \sum_{l = 0}^\infty (l+1)
{c^\dagger_l} c^{\vphantom{\dagger}}_l \ .
\end{equation}
With $D$ fermions the ground state has levels $l = 0,1,
\ldots,D-1$ filled, forming a bubble of radius $r_0 = \sqrt{2D/B}$.
The number of states per area is
\begin{equation}
\rho = \frac{D}{\pi r_0^2} = \frac{B}{2\pi}\ ,
\end{equation}
independent of $D$.
Low-lying excitations involve fermions and holes with $l$ close to $D$,
which by eq.~(\ref{radius}) are near the edge.  
The level spacing $\kappa/B$ corresponds to a massless field with velocity $v
= r_0 \kappa/B$.  This is the same velocity that one gets by balancing the
Lorentz force against that from the confining potential.

We are interested in the limit of an infinite bubble, where the edge $S^1$
becomes the real line $\R$.  Take $r_0$ to infinity while holding
$B$ and $v$ fixed, and focus on a point on the edge, say $x_a = (0, -r_0)$. 
By translation invariance we can take this point to be the origin, and in the
limit the potential linearizes, 
$V = -v B x_2$.  Then
\begin{equation}
H' = -v B x_2 = v \Pi_1 \quad (\mbox{between LLL states})\ .
\label{oneham}
\end{equation}
The last equality follows from $\Pi_1 + B x_2 = -i(D_z + D_{\bar
z})/2$, since $D_z$ $(D_{\bar z})$ gives
zero acting to the left (right).  Equivalently, it reflects the
noncommutativity in the lowest Landau level, $[x_1,x_2] = -i/B$.  The
Hamiltonian~(\ref{oneham}) describes fermions moving to the left with velocity
$v$.  The second quantized description is
\begin{equation}
{\bf H}' = -iv \int_{-\infty}^\infty dx_1\, 
\Psi^\dagger
\partial_1 {
\Psi}\ .
\end{equation}

\subsection{The $U(1)$ QHE in four dimensions}

\subsubsection{The bulk}

The most direct extension of the QHE to four dimensions is to introduce
constant
$U(1)$ magnetic fields in two independent planes,
\begin{equation} 
F_{12} = F_{34} = B\ .
\end{equation}
The Hamiltonian is
\begin{eqnarray}
H &=& -\frac{1}{2m} D_a D_a \nonumber\\
&=& \frac{1}{2m} \biggl[ -\partial_a \partial_a + \frac{B^2}{4} x_a
x_a - B (L_{12} + L_{34}) \biggr]  \label{abfor}
\end{eqnarray}
where now $a$ runs $1,\ldots,4$.
This is just two copies of the previous system.  In particular, we can
introduce two complex coordinates $z_\alpha$,
\begin{equation}
z_1 =  \frac{1}{2}(x_1 + i x_2)\ , \quad z_2 =  \frac{1}{2}(x_3 + i x_4)\ ,
\end{equation}
and the lowest Landau level consists of all states of the form 
\begin{equation}
\psi = f(z_1,z_2) \exp(-B z^\dagger{\cdot} z )\ .
\end{equation}
where $z^\dagger{\cdot} z = \OL{z_1}z_1 + 
\OL{z_2}z_2$.  The background can be written
\begin{equation}
F_{\alpha\bar\beta} = 2iB\delta_{\alpha\bar\beta}\ ,\quad F_{\alpha\beta} =
F_{\bar\alpha\bar\beta} = 0\ .
\end{equation}
In this form there is a manifest $U(2)$ symmetry,
\begin{equation}
z_{\alpha} \to M_{\alpha \beta} z_\beta
\label{u2}
\end{equation}
for any $2\times 2$ unitary matrix $M$.  There are also translational
symmetries in the four dimensions.

\subsubsection{The boundary}

The confining potential 
\begin{equation}
V = \kappa x_a x_a / 2 = {2\kappa} z^\dagger{\cdot} z
\label{confour}
\end{equation}
gives two copies of the two-dimensional system~(\ref{conf}).  For example,
\begin{equation}
\langle l^{\vphantom\prime}_1 l^{\vphantom\prime}_2 | V | l_1' l_2'\rangle
=
\frac{\kappa}{B} (l_1 + l_2 +2)
\delta_{l^{\vphantom\prime}_1 l_1'}
\delta_{l^{\vphantom\prime}_2 l_2'}\ ,
\end{equation}
where $l_1$ and $l_2$ are the eigenvalues of $L_{12}$ and $L_{34}$.  This
potential preserves the $U(2)$ symmetry~(\ref{u2}) while breaking the
translational symmetries.

Now let us go to the linearized limit,
\begin{equation}
V = u_a x_a\ .
\label{linfour}
\end{equation}
By a $U(2)$ rotation we can take $(u_1 + i u_2,u_3 + i u_4)$ to
$(0, -i v B)$ so that the confining force is in the 4-direction.  This
corresponds to looking at a point on the sphere that is tangent to the
1-2-3 plane.  Then 
\begin{equation}
H' = -v B x_4 = v P_3\quad (\mbox{between LLL states})\ .
\end{equation}
We thus have two copies of the two-dimensional system. 
The first, in the 1-2 plane, has no potential and so an
infinitely degenerate ground state.  The second, in the 3-4 plane, has a
linear potential and one-dimensional edge dynamics.  We can use the
$L_{12}$ basis for the first and the
$P_3$ basis for the second, so that there is an infinite number of
one-particle states $\psi_{l_1,p_3}$ with given momentum $p_3$.

The second-quantized description thus involves an infinite number of
fields,
\begin{equation}
{\bf H}'
= -iv \int_{-\infty}^\infty dx_3\, \sum_{l=0}^\infty \Psi_l^\dagger
\partial_3
\Psi_l\ .
\end{equation}
Here $l \equiv l_1$ is the helicity, the eigenvalue of the
rotation $L_{12}$ around the direction of motion.
Alternatively,
\begin{equation}
{\bf H}' = - i v \int  d^3x\, \Psi^\dagger(\vec x)  \partial_3 \,
\Psi(\vec x)\ ,
\end{equation}
but with the 1-2 plane noncommutative, $[x_1,x_2] = -i/B$.
The boundary theory is not truly three-dimensional, but rather
one-dimensional with an infinite number of fields.  We can understand this
in terms of the symmetries of the system.  We have noted that the
confining potential~(\ref{confour}) leaves a $U(2)$ spatial symmetry.  In
the linear limit~(\ref{linfour}) the four symmetry generators become the
translations in the 1-, 2-, and 3-directions and the rotation in the 1-2
plane.  We are missing the additional two rotational symmetries of
$\R^3$, which would rotate the 3-direction into the other two and so require
fields moving in all directions.

\subsubsection{Particle-hole states}

Although the $U(1)$ system is not truly three-dimensional, it is a useful
warmup for the $SU(2)$ system, and so we develop some of the properties of
its particle-hole states.  We focus on the two-body wavefunction
\begin{equation}
\psi(x,x') = \langle 0|  \Psi (x)  \Psi^\dagger(x') | \Sigma\rangle
\end{equation}
where $| \Sigma\rangle$ is a particle-hole state.

One basis for the particle-hole states is
\begin{equation}
\psi(x,x') = \psi^{\vphantom *}_{l_1,p_3}(x) \psi^*_{-l_1',-p_3'}(x')\ ,
\label{helbas}
\end{equation}
taking the particle and hole each to have definite 3-momentum and definite
helicity.  The total quantum numbers for the pair are then $P_3 = p_3 + p'_3$
and $L_{12} = l_1 + l_1'$.  In particular there is an infinite number of
ways to get $L_{12} = \pm 2$.  

The total particle-hole momenta are $\Pi_a = \Pi^{\rm p}_a +
\Pi^{\rm h}_a$ with $\Pi^{\rm p}_a = -i \partial_a +  A_a(x)$ and $\Pi^{\rm
h}_a = -i \partial'_a -  A_a(x')$.  Note that unlike the separate
particle and hole momenta, the total momenta commute,
$[\Pi_a,
\Pi_b] = 0$.  Thus we can take for example a basis that are eigenstates
of $\Pi_1$, $\Pi_2$,
$\Pi^{\rm p}_3$, and $\Pi^{\rm h}_3$ with respective eigenvalues $P_1$,
$P_2$, $p_3$, and $p'_3$.  One finds 
\begin{eqnarray}
\psi_{P_1,P_2,p_3,p'_3}(x,x') &\propto&
\exp\biggl\{ -B\Bigl( z^\dagger{\cdot} z + z'^\dagger{\cdot} z'
- 2 z_1 \OL{z'_1} - z_2^2 - \OL{z'_2}^2 \Bigr) 
\nonumber\\
&&\quad + i (P_1 - i P_2) z_1 
+ i (P_1 + i P_2) \OL{z'_1} + 2 i p_3 z_2 +  2 i p'_3 \OL{z'_2}\, \biggr\}\
.
\label{mombas}
\end{eqnarray}
In the 1-2 plane these are gaussian in the separation and plane waves in
the center of mass.  In the 3-4 plane they are plane waves in $x_3$ and
$x_3'$ and gaussian in $x_4$ and $x'_4$.

The states~(\ref{helbas}) and~(\ref{mombas}) are both nonseparating:
the particle and hole move in the 3-direction with fixed velocity, while
in the 1-2 plane they are confined by the magnetic force as
argued in ref.~\cite{hz1}.  The loophole in the argument given in the
introduction is that the velocity here is $v_a = v \delta_{a3}$, {\it
independent} of the momentum: bosonization is possible because the
dynamics is one-dimensional. 

To obtain a relativistic theory we should retain only states where the
momentum is proportional to the velocity.  The states with this property
are the momentum eigenstates~(\ref{mombas}) such that $P_1 = P_2 = 0$. 
Note however from their explicit form that all these states have helicity
identically zero: they are invariant under simultaneous rotation of $z_1$
and $z'_1$.  This is an obstacle to a relativistic theory
with spin.  

Refs.~\cite{hz1,hz2} identify {\it extreme dipole states (EDS),} which are
the candidate graviton states.  These have an analog in the $U(1)$ model. 
To make contact with the notation of ref.~\cite{hz2} we start with the
spherically symmetric potential~(\ref{confour}).  The EDS are
eigenstates of the
$SU(2)$ part of the unitary symmetry~(\ref{u2}).  Call this symmetry
$K_{1i}$ where $i = 1,2,3$, and the total for a particle-hole pair is $T_{1i}
= K_{1i} + K'_{1i}$.  Let the particle have total harmonic oscillator level
$n$ and the hole total level
$n'$.  The LLL states are sums of monomials of degree $n$ in $z_\alpha$ and of
degree $n'$ in $\OL{z_\beta}'$, times an invariant gaussian, so $k_1 = n/2$ and
$k'_1 = n'/2$.  Then $t_1 \geq (n-n')/2$, and the EDS are defined to saturate
this inequality, $t_1 = (n-n')/2$.  One readily finds that these states are
of the form
\begin{equation}
\psi^{\rm EDS}_m(x,x') \propto z_1^m z_2^{n - n' - m} (z'^\dagger{\cdot}
z)^{n'}
\exp\Bigl\{ -B( z^\dagger{\cdot} z + z'^\dagger{\cdot} z'
)\Bigr\}\ .
\label{eds}
\end{equation}

To make contact with the basis~(\ref{mombas}) we must expand near the boundary,
\begin{equation}
(z_1, z_2) = (\tilde z_1, \tilde z_2) + (0,-ir_0/2)\ .
\end{equation}
Also, because the
vector potential is translation-invariant only up to a gauge
transformation we must transform to
\begin{equation}
\tilde\psi = U \psi\ ,\quad U = e^{i B r_0 (x'_3 - x\vphantom{'}_3)/2} \
.
\end{equation}
This is determined by $H\{\tilde z, \partial_{\tilde z} \} = U H\{ z,
\partial_z\} U^{-1}$.  
The tilded wavefunction in the tilded coordinates is to be compared
(dropping the tildes) to the wavefunctions~(\ref{mombas}) obtained directly
near the origin.  

From the discussion in section~2.1 it follows that as $r_0 \to
\infty$,
states of fixed energy relative to the Fermi level have
\begin{equation}
n = B r_0^2 / 2 +  r_0 q\ ,\quad n' = B r_0^2 / 2
- r_0 q'
\end{equation}
with $q$ and $q'$ fixed. Taking the limit of the
states~(\ref{eds}) with this scaling gives
\begin{equation}
\tilde\psi^{\rm EDS}_m \to z_1^m \psi_{0,0,q,q'}\ .
\end{equation}
Thus for $m = 0$ the EDS is the zero-helicity plane wave state encountered
above, while for positive $m$ we obtain a non-normalizable state of
helicity $m$.  We conclude that the EDS of nonzero helicity are not good
states in the $\R^3$ limit.  We can also understand this as follows.
One finds that $U T_{1i}\{ z,
\partial_z\} U^{-1} = -i r_0 \Pi_i / 2$, so that the EDS condition
linearizes to $(\Pi_1^2 + \Pi_2^2) \psi = 0$.  The only normalizable
solutions again have $P_1 = P_2 = 0$, but multiplying by a power of $z_1$
gives a nonnormalizable solution.  Thus we can characterize the EDS with
$m \neq 0$ as states of definite helicity and definite momentum-squared,
but indefinite momentum. 
One can generalize the EDS to $t_1 = s + (n-n')/2$ with fixed $s$.  This
introduces an extra power of $\OL{z_1}'^s$ in the flat limit, allowing
negative helicities but still non-normalizable.

The energy of a particle-hole state is $E = v(n - n')/r_0 = v(q + q')
= v P_3$.  The EDS states thus have a relativistic dispersion relation
$E^2 = v^2 P^2$.  Note that the non-EDS states are all tachyonic (in the
sense of their momenta, not their velocities):
$E^2 = v^2 P_3^2 < v^2 P^2$.  This is a further obstacle to obtaining a
relativistic theory.

\sect{The $SU(2)$ QHE on $\R^4$}

\subsection{The model}

By extending to an $SU(2)$ magnetic field it is possible to obtain a
larger spatial symmetry~\cite{hz1}.  Consider the configuration
\begin{equation}
F^1_{23} = F^1_{14} = F^{2}_{31} = F^2_{24} = F^3_{12} = F^3_{34} = B\ .
\label{su2mag}
\end{equation}
In other words, $F^i_{ab} = B \eta^i_{ab}$ where
\begin{equation}
\eta^i_{ab} = \epsilon_{iab 4} + \delta_{ia} \delta_{4b}
- \delta_{ib} \delta_{4a}
\end{equation}
is the 't Hooft symbol.  Note that $a,b$ run $1,\ldots,4$
and $i,j$ run $1,\ldots,3$.

Let us analyze the symmetries of this configuration.  First use the
separation of $SO(4)$ into two commuting $SO(3)$ algebras,
\begin{eqnarray}
K_{1i}^{(0)} &=& -\frac{1}{4}\tilde\eta^i_{ab}L^{\vphantom i}_{ab}
=\frac{1}{2}(L_i + L_{4i})
\ ,\nonumber\\
K_{2i}^{(0)} &=&\frac{1}{4}\eta^i_{ab}L^{\vphantom i}_{ab}
= \frac{1}{2}(L_i -  L_{4i})\ ,
\end{eqnarray}
where
\begin{equation}
\tilde\eta^i_{ab} = -\epsilon_{iab 4} + \delta_{ia} \delta_{4b}
- \delta_{ib} \delta_{4a}
\end{equation}
is the parity-reflected 't Hooft symbol.
We follow the notation of refs.~\cite{hz1,hz2}.  We can similarly separate
the field strength
\begin{equation}
G^i_{1j} = -\frac{1}{4}\tilde\eta^j_{ab}F^{ i}_{ab}\ ,
\quad G^i_{2j} = \frac{1}{4}\eta^j_{ab}F^{ i}_{ab}\ .
\end{equation}
Then $G^i_{1j}$ is invariant under $K_{2}^{(0)}$, while it transforms as a
vector of $K_{1}^{(0)}$ on its $j$ index.  Similarly $G^i_{2j}$ is
invariant under $K_{1}^{(0)}$, while it transforms as a vector of
$K_{2}^{(0)}$ on its $j$ index.  Also, each is a vector of isospin $I$
on its $i$ index.  In this notation the configuration~(\ref{su2mag}) is
\begin{equation}
G^i_{1j} = 0 \ , \quad G^i_{2j} = B \delta^i{}_j/2\ .
\end{equation}
It follows that this is invariant under $K_1^{(0)}$ and under
simultaneous rotation by $K_2^{(0)}$ and by $I$.  Thus we
define~\cite{hz1}
\begin{equation}
K_{1i} = K_{1i}^{(0)}\ , \quad
K_{2i} = K_{2i}^{(0)} + I_{i}\ , \label{su2s}
\end{equation}
which are the symmetries of this configuration; here $I_i$ is the
$(2I+1)$-dimensional representation of $SU(2)$.  The
generators~(\ref{su2s}) form an
$SO(3)
\times SO(3) = SO(4)$ algebra, all generators of which act nontrivially on
space.  The generators $K_{2i}$ have also an action on the $SU(2)$ isospin
indices.

The actual model that we will study is slightly different from the above
but has the same symmetries.  That is, we will take the vector potential
\begin{equation}
A^i_a = -\frac{B}{2} \eta^i_{ab} x_b\ . \label{su2vect}
\end{equation}
In the corresponding field strength, 
\begin{equation}
F^i_{ab} = \partial_a A^i_b - \partial_b A^i_a 
+ \epsilon_{ijk} A^j_a A^k_b
\ , 
\end{equation}
the linear terms reproduce the earlier configuration~(\ref{su2mag}), but
the quadratic term is nontrivial and of order $x^2$.  We take the
potential to be simple, rather than the field strength, because it is this
that appears in the Hamiltonian.

The configuration~(\ref{su2vect}) is invariant under $SO(4)$ rotations but
it is clearly not translationally invariant because of the $O(x^2)$ terms
in the field strength.  However, the
confining potential that is to be added breaks these same
translation symmetries.  

Curiously, the configuration~(\ref{su2mag}), in spite of its simple
appearance, is not translationally invariant either.  That is, there is
no magnetic translation $\Pi_a$
having the property
\begin{equation}
[\Pi_a, D_b] = 0  \label{mtdef}
\end{equation}
for all $a,b$.
Here the covariant derivative is
\begin{equation}
D_a = \partial_a - i A_a^i I_i \equiv \partial_a - i {\bf A}_a
\ ,
\end{equation}
while 
\begin{equation}
\Pi_a = -i(\partial_a - i {\bf V}_a)
\end{equation}
is the combination of a translation in the $a$-direction with some
infinitesimal gauge transformation ${\bf V}_a$.  To show that there is
no such symmetry, note first that the property~(\ref{mtdef}), with
the Jacobi identity, implies
\begin{equation}
\Bigl[ [ \Pi_a,\Pi_b ], [D_\rho,D_\sigma] \Bigr] = 0
\ \ \Rightarrow [ {\bf W}_{ab} , {\bf F}_{cd} ] = 0\ .
\end{equation}
Here
\begin{equation}
{\bf F}_{cd} = { F}^i_{cd} I_i
= \partial_c {\bf A}_d - \partial_d {\bf A}_c 
- i [ {\bf A}_c, {\bf A}_d]
\end{equation}
is the
field strength in matrix notation, while ${\bf W}_{ab}$ is similarly
constructed from ${\bf V}_a$.  Since the ${\bf F}_{cd}$ span a
complete set of $SU(2)$ generators it follows that
\begin{equation}
{\bf W}_{ab} = 0\ \ \Rightarrow\ \ {\bf V}_a = g \partial_a g^{-1}
\end{equation}
for some $g(x)$ in $SU(2)$.  But then the definition~(\ref{mtdef}) implies
\begin{equation}
[g \partial_a g^{-1}, D_b ] = 0\ \ \Rightarrow\ \ 
[\partial_a, g^{-1} D_b g ] = [\partial_a, \partial_b
- i {\bf
A}^{g}_b ] = 0\
.
\end{equation}
That is, there is a gauge in which the vector potential ${\bf
A}^{g}_a$ is constant and so
\begin{equation}
{\bf F}^g_{cd} =
- i [ {\bf A}^g_c, {\bf A}^g_d]\ .
\end{equation}
Finally, let $c = 1$ and let $d$ run over $2,3,4$.  Then the
left-hand side runs over a complete set of independent $SU(2)$ generators,
while the right cannot (its trace with ${\bf A}^g_c$ always vanishes).
QED

Essentially, the naive translational invariance of the
configuration~(\ref{su2mag}) is broken by the action of parallel transport
on the isospin index. 
It is interesting to compare this with the Zhang-Hu
configuration~\cite{hz1} which has the larger symmetry $SO(5)$.  One can
think of the gauge curvature in that configuration as conspiring with the
curvature of the
$S^4$ to allow the extra symmetries to exist.  This is one reason why in
that system the gauge field strength must go to zero as the radius of the
$S^4$ goes to infinity, and so why the isospin must be taken to infinity
to get a nontrivial limit.  By keeping only $SO(4)$ symmetry from
the start it is possible to find a larger set of models on the flat $\R^4$.

However, there will ultimately be a penalty for the lack of translation
invariance.  In the usual QHE, the combination of translation invariance and
localized states implies an infinitely degenerate LLL with a uniform density
of states.  This will not be the case here, and will necessitate tking the $I
\to \infty$ limit.

\subsection{The spectrum}

The Hamiltonian for a spinless particle coupled to the vector
potential~(\ref{su2vect}) is
\begin{equation}
H = -\frac{1}{2m} D_a D_a + \frac{\kappa}{2} x_a x_a = H_1 + H_2\ ,
\label{su2ham}
\end{equation}
where $H_1$ is the oscillator Hamiltonian
\begin{equation}
H_1 = \frac{1}{2m} (-\partial_a\partial_a + m^2 \omega^2 )\ ,
\quad
m^2 \omega^2 = \frac{B^2}{4} I (I+1) + m\kappa\ ,
\end{equation}
and $H_2$ is the spin-isospin interaction
\begin{eqnarray}
H_2 &=& -\frac{B}{m} K_2^{(0)} \cdot I \nonumber\\
&=& -\frac{B}{2m} \Bigl(K_2 \cdot K_2 - I \cdot I - K_2^{(0)} \cdot
K_2^{(0)}\Bigr)\ .
\end{eqnarray}
Note that we have introduced a harmonic potential from the start, since
this entails no loss of symmetry.  There is no change of variables that
reverses the sign of $B$, and the physics will depend on the sign.

It is straightforward to diagonalize the Hamiltonian by addition of
angular momenta.  However, the reader who is interested in the $\R^3$
limit of the edge need not work through the detailed counting of states and
enumeration of cases, but may jump to the next section, since in the limit
the Hamiltonian becomes even simpler.  The only result one needs from the
remainder of this section is that in order to reach the $\R^3$ limit one
must also take $I \to \infty$.  Thus the $\R^3$ limit of our model
coincides with the $\R^3$ limit of the Zhang-Hu model.

To diagonalize $H$ consider first the oscillator part.  With $n$
excitations the oscillator energy is $E_1 = (n+2)\omega$.  The raising
operators
\begin{equation}
a_a^\dagger = -\partial_a + m\omega x_a
\end{equation}
are vectors of $SO(4)$, which can also be written as matrices
\begin{equation}
a^\dagger_{\alpha}\!^{\beta} \equiv a_4^\dagger
\delta_{\alpha}{}^{\beta} + i a_i^\dagger
(\sigma^i)_{\alpha}{}^{\beta}\ .
\end{equation}
These transform as spin-$\frac{1}{2}$ under both $K_1^{(0)}$ and
$K_2^{(0)}$; the $K_1^{(0)}$ index is written as a subscript and the
$K_2^{(0)}$ index as a superscript.  At level $n$, the product of $n$
$a_a^\dagger$'s gives an
$n$-fold symmetric tensor; by subtracting traces this  decomposes into
irreducible representations
\begin{equation}
(n) \oplus (n-2) \oplus (n-4) \oplus  \ldots \oplus \{ (1)\ {\rm or}\ (0)
\}
\end{equation}
where $(r)$ denotes the rank $r$ traceless symmetric tensor.  In terms of
the $SO(3) \times SO(3)$ quantum numbers $(k_1^{(0)}, k_2^{(0)})$, the
representation $(r)$ is $(\frac{1}{2}r,\frac{1}{2}r)$ and so at level $n$
the states are
\begin{equation}
(\half{n},\half{n}) \oplus
(\half{n}-1,\half{n}-1)\oplus
(\half{n}-2,\half{n}-2)\oplus
 \ldots \oplus
\{ 
(\half,\half)\ {\rm or}\ (0,0) \}\ .
\end{equation}
For each value $(\frac{1}{2}r,\frac{1}{2}r)$ the quantum numbers
$k^{(0)}_{1,3}$ and $k^{(0)}_{2,3}$ run independently from
$-\frac{1}{2}r$ to $+\frac{1}{2}r$.  The total dimension is
\begin{equation}
(n+1)^2 + (n-1)^2 + (n-3)^2 + \ldots + \{ 4\ {\rm or}\ 1\} =
\frac{1}{6}(n+3)(n+2)(n+1)\ .
\end{equation}
The equality of $k_1^{(0)}$ and $k_2^{(0)}$ follows from the
operator identity $K_1^{(0)} \cdot K_1^{(0)} = K_2^{(0)} \cdot
K_2^{(0)}$.  It is also evident from the explicit form of the states,
\begin{equation}
(\half{m},\half{m}) = \Bigl\{
a^\dagger_{\alpha_{(1}}\!^{\beta_1} 
a^\dagger_{\alpha_2}\!^{\beta_2} \ldots
a^\dagger_{\alpha_{m)}}\!^{\beta_m} \Bigr\}\Bigl\{
a^\dagger_{\alpha_{[m+1}}\!^{\beta_{m+1}} 
a^\dagger_{\alpha_{m+2]}}\!^{\beta_{m+2}} \Bigr\}
\ldots \Bigl\{
a^\dagger_{\alpha_{[n-1}}\!^{\beta_{n-1}}
a^\dagger_{\alpha_{n]}}\!^{\beta_{n}} \Bigr\}|0\rangle\ ,
\end{equation}
where we symmetrize the first $m$ $\alpha$ indices and antisymmetrize the
rest in pairs: the $\beta$ indices automatically have the same
symmetry.

To diagonalize $H_2$, add $K_2^{(0)}$ and $I$ to go to a basis of
definite
$k_2$.  Then
\begin{equation}
E =  (n+2)\omega - \frac{B}{2m} [k_2 (k_2+1) - I (I+1) - k_1 (k_1 + 1)
]\ .
\end{equation}
We have used $k_1 = k_1^{(0)} = k_2^{(0)}$.
States are labeled by the quantum numbers 
\begin{equation}
(n, k_1, k_{1,3}, k_2,
k_{2,3}) \label{basis}
\end{equation}
with the ranges
\begin{eqnarray}
n &\in& \{ 0,1,2,\ldots \}\ ,\nonumber\\
k_1 &\in& \{ \half{n}, \half{n}-1,\ldots,  \half\ {\rm
or}\ 0 \}\ ,\quad
k_{1,3} \in \{ k_1, k_1-1,\ldots, -k_1 \}\ ,\nonumber\\
k_2 &\in& \{ I + k_1, I + k_1-1,\ldots,  |I - k_1 |\}\ ,\quad
k_{2,3} \in \{ k_2, k_2-1,\ldots, -k_2 \}\ .
\
\end{eqnarray}

\subsection{The lowest Landau level}

Unlike the $U(1)$ theory, the physics depends on the sign of $B$.  Thus the
analysis separates into two cases.

\subsubsection{$B > 0$}

For given $k_1$, the energy is minimized by taking $k_2$ to have its
maximum value $k_1 + I$, so that
\begin{equation}
E =  (n+2)\omega - {B} k_1 I/m\ ,\quad (k_2 = k_1 + I)\ .
\end{equation}
For given $n$, this is minimized in turn by taking $k_1$ to have its
maximum value $\half n$, and so
\begin{equation}
E =  2\omega + n (\omega - {BI}/{2m}) \ ,\quad (k_1 = \half n\ ,\ k_2 =
\half n + I)\ .
\end{equation}
In order that this be independent of $n$, we must take $\omega = 
{BI}/{2m}$ and so the harmonic potential is $\kappa = - B^2 I/4m$.  In
contrast to the $U(1)$ case, we need a harmonic potential to obtain a
large degeneracy; this is due to the lack of translation invariance of the
vector potential.

The LLL states, all with $E = 2\omega = BI/m$, are then
\begin{equation}
{\bf I}:\quad (n, \half n, k_{1,3}, \half n + I, k_{2,3})\ ,\quad
n \in \{0,1,2,\ldots\}
\label{lll1}
\end{equation}
with degeneracy $(n+1) (n+2I+1)$ for given $n$.

\subsubsection{$B < 0$}

Now for given $k_1$, the energy is minimized by taking $k_2$ to have its
minimum value $|k_1 - I|$, giving
\begin{equation}
E = \left\{ \begin{array}{l} (n+2)\omega - |B| k_1 (I+1)/m
\ ,\quad (k_2 = I - k_1 \geq 0)\ ,
\\[4pt]
(n+2)\omega - |B| (k_1 + 1) I/m
\ ,\quad (k_2 = k_1 - I \geq 0)\ .
\end{array}
\right.
\end{equation}
For given $n$ and either sign of $I-k_1$, this is again minimized by
taking
$k_1$ to have its maximum value $\half n$, and so
\begin{equation}
E = \left\{ \begin{array}{l} 2 \omega + n (\omega - |B|[I+1]/2m )
\ ,\quad (k_1 = \half n\ ,\ k_2 = I-\half n \geq 0)\ ,
\\[4pt]
2 \omega - |B| I/m + n (\omega - |B| I /2m)
\ ,\quad (k_1 = \half n\ ,\ k_2 = \half n - I \geq 0)\ .
\end{array}
\right.
\end{equation}
There are now two values of $\kappa$ that give a large ground state
degeneracy. For $\kappa = B^2 (I+1)/4m$ so that $\omega = |B|[I+1]/2m$, the
states with
$n \leq 2I$ are degenerate and lie below those with $n > 2I$.
For $\kappa = - B^2 I/4m$ so that $\omega = |B|I/2m$, the states with 
$n \geq 2I$ are degenerate and lie below those with $n < 2I$.

To summarize, for
$\kappa = B^2 (I+1)/4m$ the LLL states have
$E = |B|[I+1]/m$ and quantum numbers
\begin{equation}
{\bf II}:\quad (n, \half n, k_{1,3}, I - \half n , k_{2,3})\ ,\quad
n \in \{0,1,2,\ldots , 2I\}
\label{lll2}
\end{equation}
with degeneracy $(n+1) (2I-n+1)$ for given $n$.  
For $\kappa = -B^2 I/4m$ the LLL states have $E =
|B|/m$ and quantum numbers
\begin{equation}
{\bf III}:\quad (n, \half n, k_{1,3}, \half n - I , k_{2,3})\ ,\quad
n \in \{2I,2I+1,\ldots \}
\label{lll3}
\end{equation}
with degeneracy $(n+1) (n - 2I+1)$.  

\subsubsection{Discussion}

The next step is to find the boundary theory, increasing the harmonic
potential slightly so as to confine a finite bubble of fermions, and
then taking the size of the bubble to infinity while focusing on a
point on the boundary.  We have three LLL systems to work with, labeled
I, II, and III above.  

However, none
of these allows a straightforward limiting process. Consider the mean value of
$x_a x_a = r^2$ in the LLL states.  Since the LLL states have distinct
$SO(4)$ quantum numbers, $r^2$ is diagonal in the basis~(\ref{basis}) and a
short calculation gives
\begin{equation}
r^2 = \frac{n+2}{m\omega} \quad({\rm LLL})\ .
\end{equation}
The volume of the shell between $n$ and $n+1$ is then
\begin{equation}
V = 2\pi^2 r^3 \frac{\delta r}{\delta n} \approx \frac{2\pi^2 r^2}{|B|I}\ .
\label{volume}
\end{equation}
We take $n, I \gg 1$ so that the levels are closely spaced.  The
number of states in the shell, divided by the volume $V$, is
\begin{eqnarray}
{\bf I}:&&\ \rho= \frac{|B|^2 I^3}{2\pi^2} (1 + |B| r^2/4) \nonumber\\
{\bf II}:&&\ \rho=\frac{|B|^2 I^3}{2\pi^2} (1 - |B| r^2/4) \nonumber\\
{\bf III}:&&\ \rho=\frac{|B|^2 I^3}{2\pi^2} (-1+|B| r^2/4) \ .
\label{nstates}
\end{eqnarray}
The range of $r$ is implicitly limited by the positivity of $\rho$.
In all cases $\rho$ is a nontrivial function of $r$.  This is in contrast
to the familiar Abelian case where the density is constant.  The
$r$-dependence would not be present if the LLL were translation invariant,
but we have emphasized that this invariance is absent.
If we try to make an boundary system on $\R^3$ by taking $r \to \infty$ in case
I or III, the limit is singular because the local density of states diverges as
$r^2$.  In case II we do not even have this option: the LLL has a finite
radius even in the absence of a confining potential.

Note that
the density of states is constant in cases I and II in the limited range
$r^2 \ll |B|^{-1}$.  However, in order to take $r \to \infty$ we must take
$B \to 0$, and then must also take $I \to \infty$ to get a nontrivial result.
Equivalently, $r^2 \ll |B|^{-1}$ is $n \ll I$, so $n \to \infty$ implies $I
\to\infty$.  Thus, while we are able to formulate the $SU(2)$ QHE on $\R^4$
for finite
$I$, when we attempt to reach the boundary theory on $\R^3$ we are forced to
take the same limit as in refs.~\cite{hz1,hz2}.  

In fact, our case II is very
similar to the Zhang-Hu model on $S^4$.  In both cases the LLL has a finite
number of states, and the $SO(4)$ representations are the same,
\begin{equation}
(k_1,k_2) = (\half n, I - \half n)\ ,\quad n\in \{ 0,1,\ldots,2I\}\ .
\end{equation}
The total degeneracy 
\begin{equation}
\sum_{n=0}^{2I} (n+1)(2I - n + 1) = \frac{1}{6}(2I+1)(2I+2)(2I+3)
\end{equation}
is then the same.  In the Zhang-Hu model the LLL is uniformly distributed
on $S^4$.  Roughly speaking, one can think of our case II as cutting this
open at the north pole and spreading it out to form a ball on $\R^4$.  Near
the origin of $\R^4$, corresponding to the south pole of $S^4$, the Zhang-Hu
system and ours match; this is the region of interest for reaching the
limit of flat $\R^3$.

\sect{The $I \to \infty$ limit}

\subsection{The bulk}

We have concluded that we must keep $I\gg n$ as $n \to \infty$.  
It is logical therefore to first take $I \to \infty$ at fixed $n$, and then $n
\to
\infty$.  We have been unable to avoid the problem of an infinite-dimensional
$SU(2)$ representation, but at least we can make a virtue of necessity and take
advantage of the simplifications that occur when $I \to \infty$.  Also, this
is more closely parallel to the usual QHE, where the Hamiltonian is held fixed
(aside from scaling the confining potential) as the size of the bubble is
taken to infinity.  Note that there is another limiting process as well,
taking $m \to 0$ to restrict to the LLL.  This limit commutes with $I \to
\infty$; for example, in either order the ratio $\rho/I$, where $\rho$ is the
density of LLL states, approaches the
$r$-independent value $b^2/2\pi^2$.  It is simplest
to take the limits in the order
$I\to\infty$, then $m \to 0$, and finally $n \to \infty$.

In order to obtain a nontrivial $I \to \infty$ limit of the
Hamiltonian~(\ref{su2ham}), we must hold fixed $b = BI$; in this same limit
$\kappa \to 0$ and the Hamiltonian becomes
\begin{equation}
H = \frac{1}{2m} \biggl(-\partial_a \partial_a + \frac{b^2}{4} x_a x_a
- 2b\, \vec e \cdot \vec K_{2}^{(0)} \biggr)\ .
\label{limham}
\end{equation}
Here we have defined
\begin{equation}
e_i = {I_i}\Big/{\sqrt{I(I+1)}}\ .
\end{equation}
Since
\begin{equation}
[ e_i,  e_j] =  {i} \epsilon_{ijk}  e_k\Big/{\sqrt{I(I+1)}}\
,\quad \vec e
\cdot \vec e = 1\ ,
\end{equation}
$\vec e$ becomes a {\it classical} unit vector as $I \to \infty$.  

The Hamiltonian~(\ref{limham}) is the same as the Abelian
Hamiltonian~(\ref{abfor}), with the replacements
\begin{equation}
B \to b\ ,\quad L_{12} + L_{34} \to 2 \vec e \cdot \vec K_2^{(0)}
= e_i (L_i -  L_{4i})\ .
\end{equation}
In particular, for $\vec  e = (0,0,1)$, $2 \vec e \cdot \vec K_2^{(0)} =
L_{12} + L_{34}$ and the Hamiltonians are identical.  Thus we have a simple
interpretation of this system in the $I \to \infty$ limit: it is an infinite
number of copies of the $U(1)$ quantum Hall system on $\R^4$, with the
spatial orientation of the magnetic field indexed by the unit vector $\vec 
e$.  Note that in the limit translation invariance on $\R^4$ is restored.

The LLL then consists of states with the appropriate analyticity 
\begin{equation}
\psi(\vec  e, x) = 
f(\vec e, z^1,z^2) e^{-b x_a x_a/4}\ ,
\end{equation}
where now the coordinates $z$ have an implicit dependence on $\vec e$,
\begin{equation}
z_1 = (u_i+ i v_i) x_i\ ,\quad z_2 = e_i x_i + i x_4\ .
\end{equation}
Here $(\vec e,\vec u,\vec v)$ form an orthonormal frame in three dimensions.
One can see this by rotating to a frame where $\vec e = (0,0,1)$, where it
reduces to the earlier
$U(1)$ analysis.\footnote
{Since the space of complex structures on $\R^4$ is part of the twistor
construction, one could say that we are now considering a Fermi liquid on
twistor space.}
One can then verify that
\begin{equation}
H = \frac{b}{m} - \frac{1}{2m} D_{\alpha} D_{\bar \alpha}
\ ,\quad D_{\alpha} = \partial_{\alpha} - b \OL{ z_\alpha}\ ,\quad
D_{\bar\alpha} = \partial_{\bar\alpha} + b z_\alpha\ .
\end{equation}

\subsection{The boundary}

As in the $U(1)$ case, the $r_0 \to \infty$ limit is equivalent to linearizing
around the origin, introducing a potential $V = -vb x_4$.  Between
LLL states this becomes
\begin{equation}
H' = v e_i P_i\ .
\end{equation}
Again, this is an infinite collection of $U(1)$ systems, with all possible
spatial orientations: the velocity of the boundary excitations is $v\vec e$.
In second-quantized form one can write for example
\begin{equation}
{\bf H}' = - i v \int  d^2e\, d^3x\, \Psi^\dagger(\vec e,\vec x) \,\vec e \cdot
\vec \partial \,
\Psi(\vec e,\vec x)\ ,
\end{equation}
but where the space is noncommutative in the directions orthogonal to $\vec e$,
$[x_i,x_j] = -i \epsilon_{ijk} e_k/b$.

As has been noted in various places, one can think of the $I \to \infty$ limit
as a six-dimensional system with a five-dimensional boundary, elevating $\vec
e$ to a coordinate.  The space is then
$\R^4 \times S^2$, and its boundary is $\R^3 \times S^2$.  However, the
boundary dynamics is still one-dimensional.  The velocity is independent of the
momentum --- it depends only on the position on $S^2$, and is tangent to
$\R^3$.  

For particle-hole states to have a finite value of $T_{2i}
= K_{2i} + K'_{2i}$ as $I \to \infty$, it is necessary to take $\vec e^{\,\rm
p} = - \vec e^{\,\rm h} \equiv \vec e$.  A basis of such states, analogous to
the plane wave basis (\ref{mombas}), would then be
\begin{equation}
| \vec e, \vec P_\perp, \vec e \cdot \vec p,  \vec e \cdot \vec p\,' \rangle
\end{equation}
where $\perp$ denotes the two dimensions orthogonal to $\vec e$; one should
note that $\vec e \cdot \vec p$ and $\vec e \cdot \vec p\,'$ are always
positive.  The
$T_1$ eigenstates are obtained as in the $U(1)$ case, while the $T_2$
eigenstates correspond to appropriate superpositions of different values of
$\vec e$, since
$T_2$ rotates $\vec e$.

From the point of view of obtaining a relativistic theory with spin, the same
problems as discussed in section 2.2.3 for the $U(1)$ case arise here. 
To obtain a relativistic theory we need in some way to truncate the
one-particle spectrum to states in which
$\vec P$ is parallel to $\vec e$.\footnote{The states that must be removed 
were termed `incoherent fermionic excitations' in
ref.~\cite{hz1}.}  
However, the only such states have
zero helicity.  The extreme dipole states of nonzero helicity are
nonnormalizable.  The states with  $\vec P$ not parallel to $\vec e$ are all
tachyonic, not in their velocities but in the sense that
$P^2 > E^2/v^2$.  Since the energy of a state is $E = ve \cdot P$, the states
with $\vec e  \parallel \vec P$ are actually the {\it highest} energy states
with given
$\vec P$.

Because of the effective one-dimensionality of the edge theory it is likely
that one can solve various four-fermion interactions by means of bosonization, 
though the $I \to \infty$ limit is somewhat subtle because $\delta(0)$ appears
in various expressions, from the $\vec e$ dependence.  
For now we just note
that the most obvious effect of interactions is to allow the relativistic
states with $E = vP$ to decay to tachyonic states with $E < vP$, which would
be a problem for obtaining a relativistic theory.

\section{Discussion}

We first summarize our conclusions.  On $\R^4$ we have formulated the $U(1)$
and $SU(2)$ quantum Hall systems, with arbitrary $SU(2)$ isospin $I$.  In
the former case the boundary theory is effectively one-dimensional.  In the
latter case it is necessary to take $I \to \infty$ in order to obtain a
boundary theory, and the result is essentially an infinite collection of
one-dimensional theories.

As claimed in refs.~\cite{hz1,hz2}, even in the free theory there are
localized gapless particle-hole excitations with arbitrary helicity.  Taking
the flat limit as we have done clarifies the nature of these states.  We have
noted some specific difficulties with obtaining a relativistic theory --- the
absence of nonzero-helicity states with $\vec v
\parallel \vec P$, and the existence of tachyonic states.  However, independent
of the relativistic application, the QHE on
$\R^4$ is a rich and interesting system.  We believe that for analyzing
any local issues the limiting form that we have obtained in section~4 is the
appropriate starting point.  In particular it will be possible to solve 
certain four-fermion interactions.

We now discuss some general aspects of the emergence of gravity from
nongravitational field theories, aside from the specific details noted above. 
Let us suppose that it is possible to add interactions to the Zhang-Hu model in
such a way that the low energy fixed point becomes Poincar\'e invariant;
likely this would require a certain degree of fine tuning.  Then as noted in
ref.~\cite{hz1}, Weinberg's theorem~\cite{weinberg} would require that the low
energy interactions of massless helicity-two states take
the form of general relativity,
if these states are present and if their interactions are nontrivial  at zero
momentum transfer. The Fierz-Pauli theorem~\cite{paulifierz} (regarding the
impossibility of coupling massless higher-spin states to conserved currents)
would then require that the states of helicity greater than two decouple.

However, under the same conditions the Weinberg-Witten theorem~\cite{weinwit}
would require that the helicity-two states actually be absent from the low
energy spectrum.  The conditions for the Weinberg-Witten theorem are quite
general --- Poincar\'e invariance and the existence of a conserved
energy-momentum tensor --- so it is difficult to see how the theorems of
ref.~\cite{weinberg} could operate without the Weinberg-Witten theorem as
well.  (Note that the energy-momentum tensor in four spatial dimensions
reduces to an energy-momentum tensor in the three-dimensional boundary theory
by integrating over $x_4$.)  Thus it appears that an interacting theory of
gravity cannot arise in this way.

One can perhaps understand this heuristically as follows.  An important
feature of gravity is that there are no local observables: to say where a
measurement is made one must specify a process of parallel transport.  This
is an essential feature of general relativity.  The Zhang-Hu model, like any
ordinary nongravitational quantum field theory, does have local 
observables. 
This would be evaded if all local operators decoupled from the low energy
physics,\footnote{This possibility was also noted by C. Johnson.} but this is
not possible for the energy momentum tensor which must have a nonzero
expectation value in any state of nonzero energy.  From this point of
view it might make more sense to look for a theory of quantum gravity in the
zero energy states of the LLL without confining potential, rather than the
edge states with the potential.  Note however the complete change of
interpretation: time is no longer associated with Hamiltonian evolution, 
rather it must emerge `holographically' from correlations in the
states.\footnote{A more sophisticated obstacle to emergent gravity, pointed
out by S. Shenker, is the holographic principle.  There is strong reason to
believe that in quantum gravity the maximum entropy in a given volume is
proportional to the surface area.  If there is an underlying nongravitational
QFT one expects the entropy to be proportional to the volume.}

In perturbative string
theory one invokes Weinberg's theorem to predict that the low energy amplitudes
will be those of general relativity, and this is borne out by explicit
calculation~\cite{scherk}.  This does not conflict with the Weinberg-Witten
theorem because string theory has no local observables --- Weinberg's theorem
uses only properties of the S-matrix,\footnote{There is also the assumption
that the low energy perturbation theory can be generated by a local
Hamiltonian, which is true in general relativity.  An earlier
paper~\cite{earlier} obtains somewhat weaker results using only properties of
the S matrix.} whereas the Weinberg-Witten theorem assumes existence
of an energy-momentum tensor.

There is in fact a well-known example of emergent gravity: the AdS/CFT
duality~\cite{malda}.  On the CFT side there is a supersymmetric gauge theory
without gravity, but at large $N$ and large 't Hooft coupling the effective
description is in terms of quantum gravity, string theory actually.  The
important point is that not only does gravity emerge, but spacetime as well. 
Only the boundary of the gravitational theory is locally realized in the gauge
theory, so there are no local bulk observables.  The local observables of the
gauge theory become boundary data in the gravitational theory~\cite{GKPW}.
Note that the bulk diffeomorphism invariance is invisible in the gauge
theory; the $SU(N)$ gauge invariance is a different gauge symmetry, which
acts as a local internal symmetry, not a local spacetime symmetry, on the
boundary.

This emergence of diffeomorphism invariance
from `nothing' is analogous to what happens in the various examples of the
emergence of gauge symmetries: in coset field theories~\cite{coset}, in lattice
models~\cite{lattice}, and in the magnetic duals to supersymmetric gauge
theories~\cite{duals}.  The essential point is that gauge symmetry and
diffeomorphism invariance are just redundancies of description.  In the
examples where they emerge, one begins with nonredundant variables and
discovers that redundant variables are needed to give a local description of
the long-distance physics.  In general relativity, the spacetime
coordinates are themselves part of the redundant description.  Thus it appears
that, as in the AdS/CFT example, the emergence of general relativity requires
the emergence of spacetime itself.

\subsection*{Acknowledgments}

We would like to thank D. Gross, C. Johnson, D. Karabali, J. Kuti, S. Shenker,
and S.-C. Zhang for discussions and comments.  This work was supported by
National Science Foundation grants PHY99-07949 and PHY00-98395, and by the
Danish Research Agency.

\end{document}